# Analog electromagnetically induced transparency for circularly polarized wave using three dimensional chiral metamaterials

HAI LIN[1,*], DONG YANG[1], SONG HAN[2], YANGJIE LIU[3], AND HELIN YANG[1]

[1]*College of Physical Science and Technology, Central China Normal University, 152 Luoyu Road, Wuhan, 430079, China*
[2] *Division of Physics and Applied Physics, School of Physical and Mathematical Sciences, Nanyang Technological University, Singapore 637371, Singapore*
[3] *Antennas Group, School of Electric Engineering and Computer Science, Queen Mary University of London, 356 Engineering Building, Mile End, London, E1 4NS UK*
*\*linhai@mail.ccnu.edu.cn*

**Abstract:** In this paper, we theoretically and experimentally demonstrate a three dimensional metamaterial that can motivate electromagnetic induced transparency (EIT) by using circular polarized wave as stimulations. The unit cell consists of a pair of metallic strips printed on both sides of the printed circuit board (PCB), where a conductive cylinder junction is used to connect the metal strips by drilling a hole inside the substrate. When a right circularly polarized wave is incident, destructive interference is excited between meta-atoms of the 3D structure, the transmission spectrum demonstrates a sharp transparency window. A coupled oscillator model and an electrical equivalent circuit model are applied to quantitatively and qualitatively analyze the coupling mechanism in the EIT-like metamaterial. Analysis in detail shows the EIT window's amplitude and frequency are modulated by changing the degree of symmetry breaking. The proposed metamaterial may achieve potential applications in developing chiral slow light devices.

## 1. Introduction

Metamaterials are artificial materials that have fascinating properties [1]. Through carefully designing the structures and geometries of subwavelength unit cells, metamaterials have the ability to manipulate the propagation behavior of EM waves, including its amplitude, phase

[2, 3], polarization [4, 5] and linear/angular momentum [6]. Many exceptional phenomena such as subwavelength imaging, hologram, beam shaper, invisible cloaks [7-9] have been demonstrated. Their applications have been expanded from the microwave band to the THz, IR and visible frequencies [10]. The attractive properties of various metamaterials rely on the interaction between incident EM waves and the meta-atoms. Due to the wave-particle duality, the underlying mechanism of coupling and interference between meta-atoms may also link the classic picture to quantum picture. In recent years, by means of manipulating the interaction between light and meta-atoms, there are lots of researches focused on using metamaterial to mimic the quantum phenomena such as Electromagnetic induced transparency (EIT), Fano resonance [11], spin-orbit interaction [12], Purcell effect [13], and so on. Among these phenomena, EIT has been considered as a possible means to realize slow light for future sensing and communication devices [14]. In most of the designs, the EIT response is implemented in the metamaterial by taking advantage of meta-atom's symmetry breaking [15]. The underlying mechanism of EIT in metamaterial system is interpreted by hybridization between the so called "dark mode" meta-atom and "bright mode" meta-atom [16]. Due to the excitation of the incident linear polarization light, subwavelength electric/magnetic dipolar or multipolar are induced in the meta-atoms. Destructive interference between the dipole/multipole will create a single transmission window that arises from an opaque background. This EIT-like spectra in metamaterial can perfectly mimic a three-level quantum system while the group velocity of incident light in these metamaterials will be greatly reduced or even stopped [17, 18]. More complex dual EIT metamaterial is also studied in [19, 20] to analog a four-level tripod quantum system.

In majority of the previous works, linearly polarized EM waves are used to excite metamaterial-based EIT. As a result, slow-light effects are limited to linearly polarized waves [21, 22]. However, circularly polarized (CP) wave is also widely used in optical devices and experiments. In recent years, different kinds of chiral metamaterials have been proposed to manipulate handedness waves, various phenomenon such as negative index [23], asymmetry transmission [24], Berry phase [25], polarization rotation [26] for CP waves have been implemented. The realization of slowing down CP light may provide the possibility to construct handedness dependent optical storage and communication devices. However to the author's knowledge, the EIT-like metamaterial for circular polarization wave is not thoroughly studied. Only very recently, the authors demonstrate the EIT effect of circularly polarized waves using Born-Kuhn type resonators mixed with split ring resonators [27]. Meanwhile, most EIT-metamaterials make use of symmetry breaking "meta-atoms" on a planar to engineer the "dark mode" and "bright mode" interference [28, 29]. These works give a clue that handedness sensitive EIT of CP light can be achieved in chiral metamaterials with symmetry breaking. Also recently, in our previous work [30, 31], a unique three-dimensional design was proposed to implement polarization-dependent EIT for linear polarization wave. The 3D structure brings in more design freedom for metamaterial excogitation. Inspired by these valuable works, we use LCP/RCP electromagnetic waves to excite EIT-like response in symmetry breaking 3D chiral metamaterial. The proposed structure shows circular dichroism (CD) and optical activity as conventional chiral metamaterials [32]. A typical EIT-like transparency window can be achieved under RCP incidence while it vanished at LCP incidence. The EIT response can be tailored through changing the asymmetry degree of the unit cells. The blueshift of EIT window can be explained by employing an equivalent circuit model. Both numericaly simulated and experimentally measured results verified these classic analog of EIT-like and related Fano-like resonance spectra. An analytical coupled oscillator model is applied to reproduce the transparency spectra that enhance the understanding of the underlying mechanism. Compared with previous work [27], the proposed structure provided an alternative simple and efficient solution to tailor the EIT-like phenomenon in metamaterial for CP waves.

## 2. Design, Model and Simulation

The proposed 3-D metamaterial with a unit cell is shown in Fig. 1(a), where the meta-atom has a continuous metallic strip in both the top and the bottom side. In order to connect the top and bottom arms, one vertical metallic cylindrical junction was fabricated to pierce through dielectric layer to form a three-dimensional structure. The bottom strip is parallel to the x-axis while the top strip is rotated with an angle $\theta = 15°$ from the x axis. The whole length of both strips is 8.2mm. In order to break the symmetry of the structure, the copper via hole is not placed at the center of each metal strip. The distance between the left end of the top side strip and the via hole's center is a=3.5*mm* while for the case of the bottom strip is c=4.7*mm*. The radius of the hole is r=0.7*mm*, and the thickness of the substrate is 0.81mm which is made up with Rogers RO4003 material with a relative permittivity 3.38.

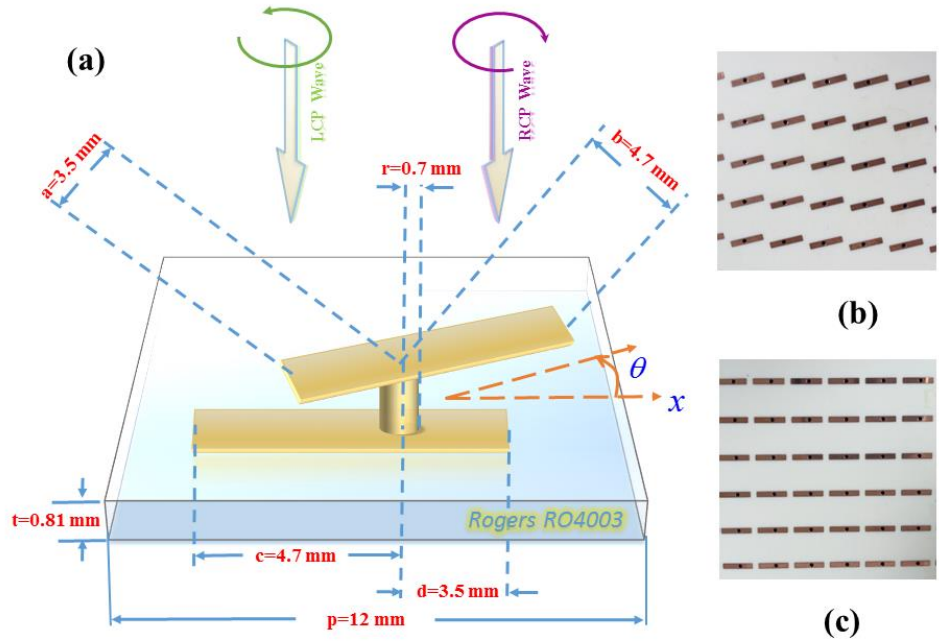

Fig.1(a) Schematic of the EIT-like metamaterial under circular polarization wave incidence，(b) top panel photograph of the fabricated sample , (c) bottom panel photograph of the fabricated sample.

The full-wave numerical simulation software CST microwave studio was employed to analyze the spectral response of our proposed design. Open boundary conditions were set along the light propagating direction (z-direction), and unit cell boundary conditions were applied at the x-y plane. The input and output Floquet ports are excited with circular polarized wave. After optimizing the parameters in simulations, experimental samples with overall dimension of 180×180 mm$^2$ were fabricated using the printed circuit board (PCB) technique, whose images of top and bottom panel are shown in Fig. 1(b) and Fig.1(c). The spectral responses were measured by circular polarized antennas used as emitter and receiver respectively. A vector network analyzer (Agilent N8362B) was employed to calibrate, store and retrieve data.

Our specific design breaks the symmetry of the 3D volumetric meta-molecular, the proposed structure has no mirror symmetry plane which can be regarded as 3D chiral metamaterial. The chirality in chiral metamaterials offers great possibilities in the control of light polarizations, optical activity and circular dichroism which can exceed the effects obtained in natural chiral materials. In this work, the chirality of the proposed metamaterial provide distinct transmission spectral for left- and right handed circularly polarized incident waves as shown in the Fig.2(a) and Fig.2(b), where $|T_{RR}|$ ($|T_{LL}|$) is defined as the amplitude ratio between the RCP (LCP) transmitted through the metamaterial and the RCP (LCP) incident wave. An absolute circular dichroism (CD) (defined as $|T_{RR}| - |T_{LL}|$ is observed in the frequency band from 10.7GHz to 13.9GHz. Meanwhile, as right handed CP wave incident from the top of the metamaterial, a sharp transparency window with the transmission peak at 11.70 GHz is switched on while it disappears under the incidence of LCP wave. It can be observed that the measurement results agree well with the simulation results.

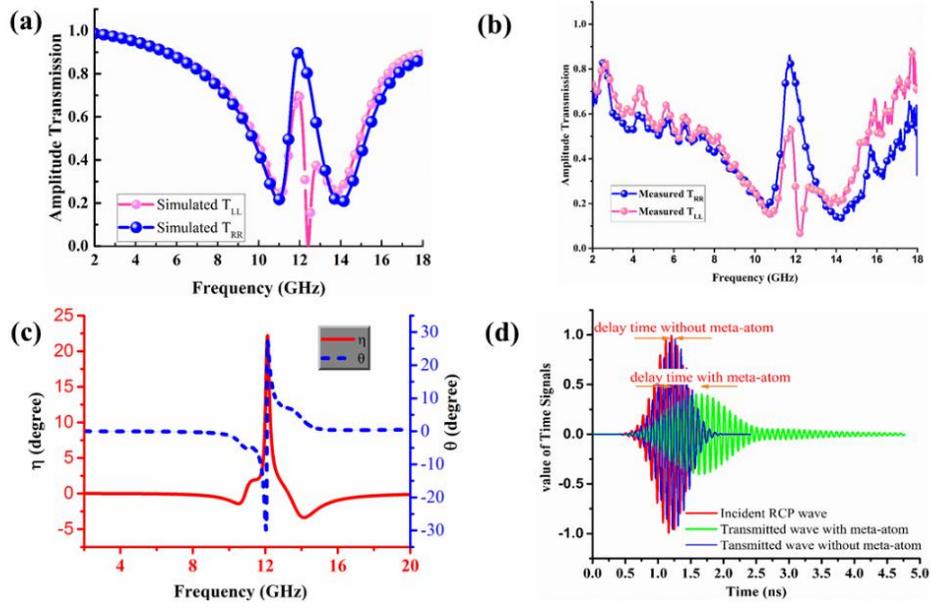

Fig.2 （a-b）Simulated and measured transmission spectra of EIT-like metamaterial under RCP and LCP incidence; (C) Simulated linear polarization rotation angle and ellipticity angle of the transmitted wave; (d) simulated time-domain pulse before and after transmitting the blank substrate, the EIT-like metamaterial sample, respectively

To analyze the chirality of the handedness EIT system, the optical activity of chiral medium is characterized by the rotation of polarization plane of a linearly polarized light as it passes through the chiral medium. It is defined as polarization rotation angle of elliptically polarized light: $\theta = [\arg(T_{RR}) - \arg(T_{LL})]/2$. Another chiral property is circular dichromism which characterizes the difference between the transmissions of two CP polarizations. A mathematical definition of CD is $\eta = 0.5 \cdot \tan^{-1}[(|T_{RR}|^2 - |T_{LL}|^2)/(|T_{RR}|^2 + |T_{LL}|^2)]$. For the proposed chiral metamaterial. An optical activity of $\pm 30°$ is observed around 12GHz as shown in Fig.2 (c), the CD property can also be observed near the transparency window.

The most important feature of EIT-like metamaterial is slow wave propagation. To demonstrate this property in the proposed design, we simulate a RCP wave carrying Gaussian shape pulse that is transmitted normally through the infinite monolayer of the metamaterial. The Gaussian pulse is centered at 11.7GHz with a 1.5GHz bandwidth. In simulation the distance between the source plane and the transmission probe plane is 10*mm*. From Fig.2 (d), it shows that the peak of the incident Gaussian pulse appears at 1.2$ns$, when the propose meta-atom (metallic part of the unit cell) exists, the peak of the transmitted pulse emerges at 1.6685$ns$. If we remove the meta-atom, the peak of the transmitted pulse shows at 1.26$ns$. This means the transmitted pulse experience a much longer delay time. Due to the strong dispersion characteristics of the metamaterial, the transmitted pulse through the metamaterial becomes much wider than the incident pulse.

In order to understanding the resonance and coupling mechanism between the meta-atoms, we plot the simulated surface current distributions under LCP/RCP light incidence in Fig.3 respectively. The frequencies are chosen as two transmission dip and transparency peak frequency for RCP incidence: 10.77GHz, 13.85GHz and 11.70GHz. For both RCP/LCP incidence case, at the lower frequency of the transmission dips, the surface currents run on the metallic structure following two loop paths due to the existence of the metal via as shown in Fig.3 (a) and Fig.3 (b). The role of each twisted loop current plays equivalent to a magnetic dipole. The direction of both magnetic dipole are antiparallel. Since the current intensity on the left loop and the right loop are comparable, the interaction between the magnetic dipoles can be regarded as strong constructive interference. At another transmission dip frequency 13.85GHz, the resonance is caused by the interaction between a twisted electric dipole and two antiparallel electric dipoles as shown in Fig.4(e) and Fig.4(f). Thus the system can be viewed to work in a symmetric mode. These working modes are quite similar to that described in [14]. However, in [14] the currents flow on a planar 2D unit cell and the stimulation wave are linear polarized.

The transparency peaks at 11.70GHz for the RCP incidence case can also be well explained using the surface current distribution. In Fig.3(d), it shows that the induced circular currents on the left and right part of the proposed structure are still antiparallel to each other. But differs from that shown in Fig.3 (b), the current density on the right path is much weaker than that of the left one. This can be regarded as an evidence of the destructive interference. On the other hand, when 11.7GHz LCP wave incidents, the loop currents runs parallel to each other, their strength are also comparable as shown in Fig.3(c), which means both twisted asymmetry U-shape resonator works at quasi-bright mode, thus, typical EIT response vanished in this case.

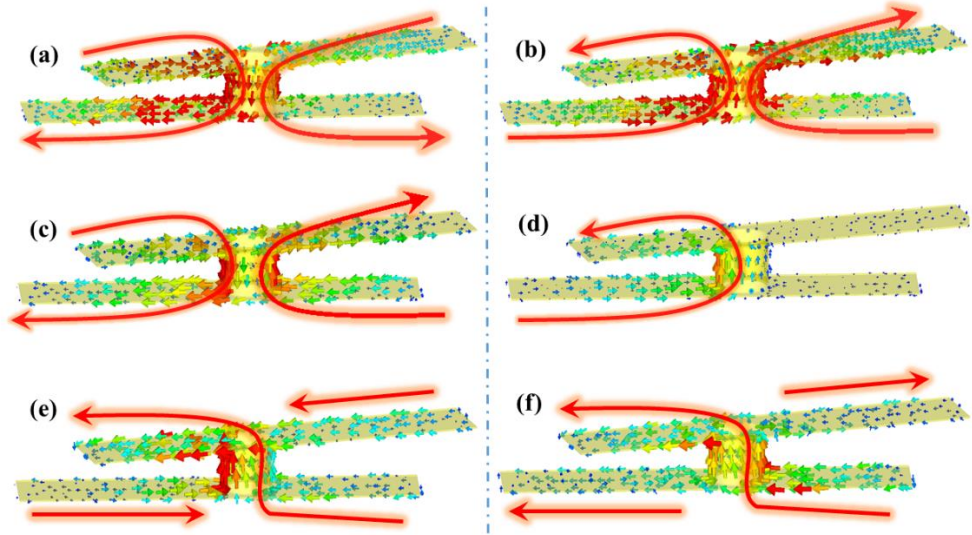

Fig.3 (a), (c) and (e) surface current distribution under LCP illumination at 10.77GHz, 11.7GHz and 13.85GHz respectively; (b), (d) and (f) surface current distribution under RCP illumination at 10.7GHz, 11.7GHz and 13.85GHz respectively

The current distributions plots illustrate that the loop currents flow along twisted-U shape paths. The coupling mechanism of the two twisted U shape resonator might have an electric analog using RLC equivalent circuit. An electrical analogy model [33] for the EIT-like phenomenon in metamaterial is considered as plotted in Fig.4 (b). The resonance atom is modeled using resonant circuit formed by coupled RLC circuit. The resistance, inductance, and capacitance of each loop are represented by $R_i$, $L_i$, $C_i$, respectively. The coupling capacitor C models the coupling between the two loop shape meta-atoms. The induced transparency is investigated by analyzing the power transfer from the voltage source to the resonant circuit $R_2$ $L_2C_2$. When geometric parameters of the structure change, the value of lumped elements in the electrical analogy model will change accordingly as well as the amplitude and frequency of transparency window. This intuitive analogy model inspire us to study tunable EIT-like phenomena based on the proposed structure which will be discussed in the subsequent section of this article.

A known theoretical model for quantitatively interpreting the EIT phenomenon is the two-oscillator model [34]. The interaction between both oscillator (bright ($x_1$) and quasi-dark ($x_2$)) with the incoming electric field $E = E_0 e^{i\omega t}$ can be quantitatively described in the following differential equations:

$$\ddot{x}_1(t) + \gamma_1 \dot{x}_1(t) + \omega_0^2 x_1(t) + \Omega x_2(t) = gE \tag{1}$$

$$\ddot{x}_2(t) + \gamma_2 \dot{x}_2(t) + (\omega_0 + \delta) x_2(t) + \Omega x_1(t) = 0 \tag{2}$$

The parameters $(\gamma_1, \gamma_2)$ and $(x_1, x_2)$ are the damping and the resonant amplitude of two oscillators, respectively. The parameter $g$ represents the coupling strength between bright mode oscillator and the external field. The coupling between the two oscillators is described by the coupling strength of $\Omega$. A detuning factor $\delta$ is introduced to represents the frequency difference between the transparency frequency and the resonance frequency of intrinsic oscillators. After solving the above coupled equations (1) and (2) with the displacements vectors expressed as $x_n = c_n e^{i\omega t} (n = 1, 2)$, and using the approximation of $\omega_1^2 - \omega^2 \approx -2\omega_1(\omega - \omega_1)$ [28], the transmission as a function of frequency can be expressed as:

$$T = 1 - \mathrm{Re} \frac{ig^2(\omega - \omega_0 - \delta + i\gamma_2/2)}{(\omega - \omega_0 + i\gamma_1/2)(\omega - \omega_0 - \delta + i\gamma_2/2) - \Omega^2/4} \tag{3}$$

Here we use the scattering parameters of an electric current sheet and the relation of $T=1-R$ [35]. Using the Curve fitting toolbox in Matlab$^{TM}$, We obtained following fitting parameters: $\gamma_1 = 4.939$, $\gamma_2 = 0.2403$, $\Omega = 2.819$, $\delta = 0.4626$, $\omega_0 = 12.55$. The analytically fitted curve using equation 3 is in good agreement with the numerical simulation results as shown in Fig. 4(a), which reflect the validity of the oscillator model. Moreover, the two parameters $\gamma_1$ and $\gamma_2$ are of different orders of magnitude, this is because the light-matter coupling extent between the incident RCP light and the two resonator are quite different.

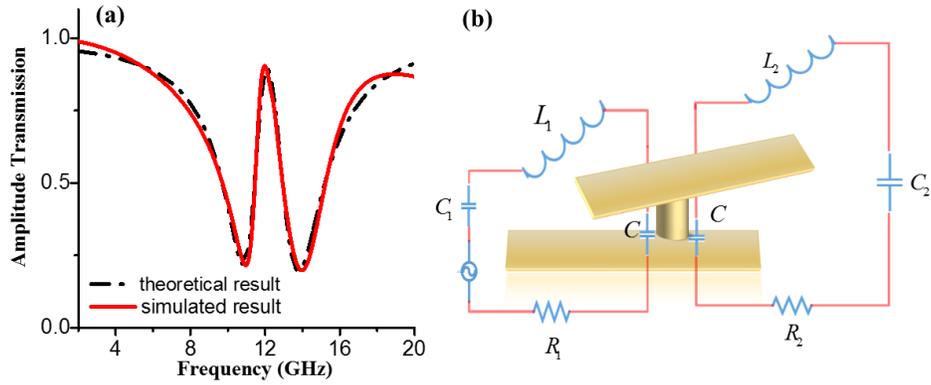

Fig.4. (a) Comparison between the simulated transmission spectra and the analytically calculated spectra based on two particle model. (b) An electric circuit analogy model for the 3D chiral metamaterial

## 3. Discussions

In a lot of previous metamaterial designs, coupled asymmetry split-ring resonators have been proposed to realize EIT-like narrow transmission resonances for linear polarized incident wave at different wavelength. To study the effect of symmetry breaking in our design, we first plot the transmission spectrum and simulated current distributions of two symmetric 3D designs which are shown in the insets of Fig. 5 separately. In these cases, the metallic vias

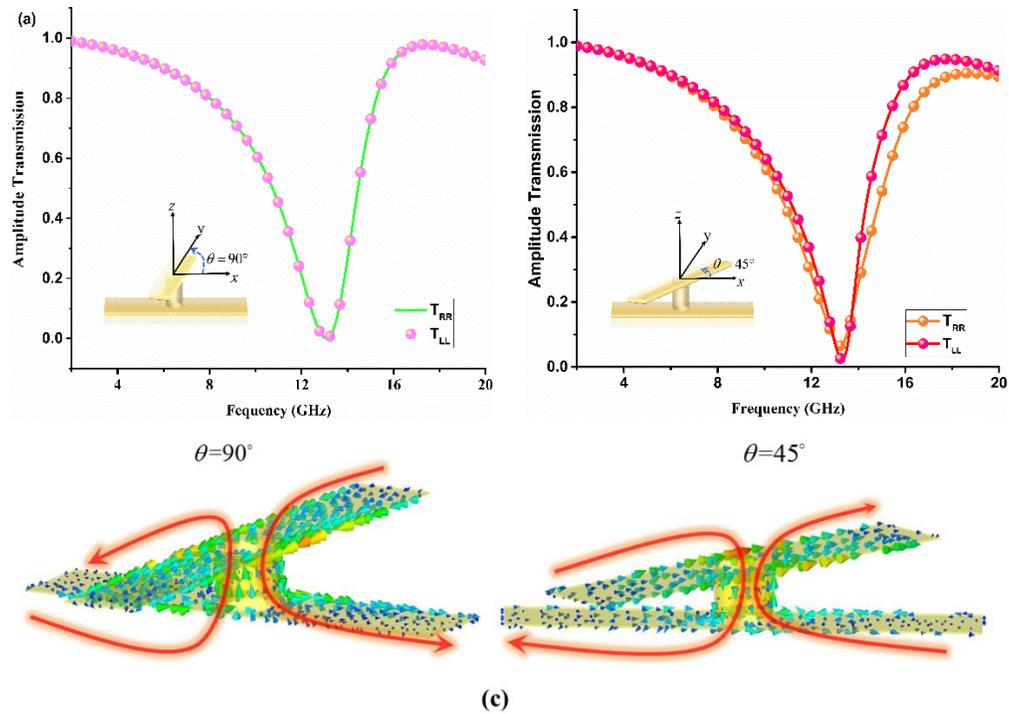

Fig.5 (a) (b)Simulated Transmission spectra of two symmetry 3D structure, the inset shows its geometric parameter, respectively; (c) surface current distribution around the resonance frequency 13.07GHz of the structures are drilled at the center of the strips on both sides, the rotation angle $\theta$ is set to be $90°$ and $45°$ separately. When $\theta = 90°$, the transmission coefficient $|T_{RR}|$ is equal to $|T_{LL}|$ since the structure is symmetric. When $\theta = 45°$, the structure is in C4 symmetry shape, thus the transmission coefficients $|T_{RR}|$ is slightly different to that of $|T_{LL}|$. The transmission dips share the same frequency whatever the circular polarization is. From the surface current distribution plot in Fig.6 (c), it can be clearly distinguished that the surface currents flow along two twist loop path in both case. Different from that in the proposed EIT-metamaterial, the surface current loops on the metallic structure introduce two magnetic dipoles which are in phase and nearly equal strength and therefore result in constructive interference.

The degree of asymmetry of our 3D meta-atoms can be tailored through changing the rotation angle $\theta$ as well as the position of the through hole. To understand the influence of different geometry parameters, the simulated amplitude transmission of the proposed design with different rotation angle $\theta$ are presented in Figure.6 (a) respectively. It shows that the amplitude of the transmission peak gradually increase with the decrease of the rotation angle ranging from $60°$ to $30°$ under the incidence of RCP light. The frequency of the transmission peak slightly increase with the increase of the $\theta$. These spectral features might be explained by the equivalent circuit model of the EIT metamaterial. As the rotation angle increase, the gap size of each twisted-U shape resonator increase which will result in the decrease of capacitance, since the resonance frequency is decided by $f = \dfrac{1}{2\pi\sqrt{LC}}$, the transmission peak thus experience a blueshift. In addition, the increased rotation angle also increase the distance between the two magnetic dipoles, and further decrease the coupling strength along with the transparency peak intensity. To study the influence of the through hole's position, we keep the rotation angle $\theta$ fixed to 15 degree and place the through hole at different location. The transmission responses are listed in Fig.6 (b). It can be observed that the line shape can be engineered from Fano shape to EIT shape with different asymmetry degree. Since the rotation angle remain constant, the current loop path's direction have not changed. Via's position only changes the coupling strength between the twisted U shape resonators. The Fano peak's frequency remain nearly unchanged while its amplitude can be modulated due to the changing of via position. Similar conclusion can also be drawn from the equivalent circuit model depicted in Fig.4 (b).

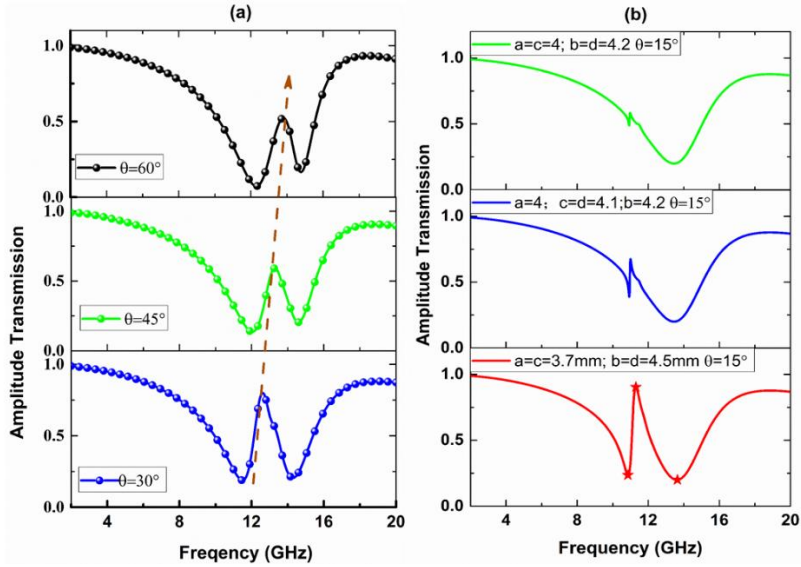

Fig.6 (a) Simulated transmission spectra with $\theta$ varying from $30°$ to $60°$ (b) Simulated transmission spectra with varying the place of the through vias while fix the rotation angle $\theta$

At last, The EIT-like metamaterials can be use as refractive-index sensor. The sensing ability of our design is depicted in Fig.7. One can clearly observe a shift of the transmission peak to lower frequency due to the change of background permittivity. The sensor may provide wide usage in environmental, chemical and other practical application areas.

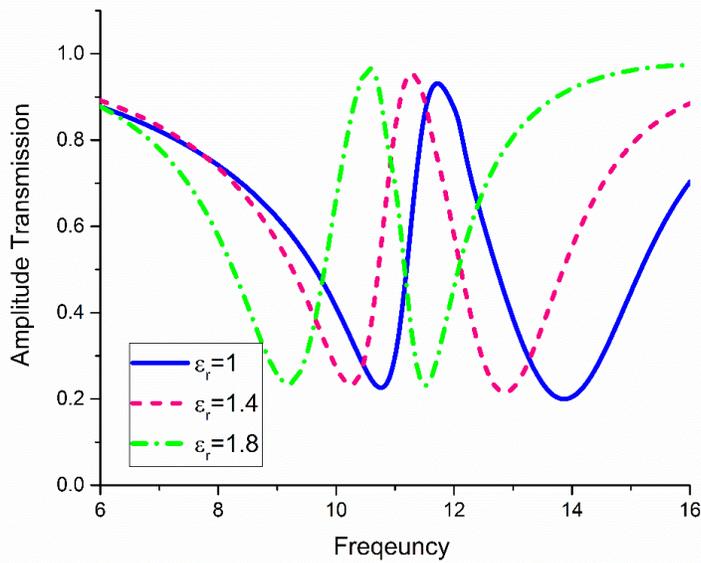

Fig.7 Simulated resonance shift of the EIT-like metamaterial with respect to the change in the relative permittivity of the background medium.

## 4. Conclusion

In summary, we have numerically and experimentally verified a novel three dimensional metamaterial structures that support classical analog of electromagnetically induced transparency and slow light behavior for circular polarized wave. The unique spectral features rely on the introduction of symmetry breaking into chiral metamaterials, which provide more resonance modes for the structure, such as constructive asymmetric mode, destructive asymmetric mode and symmetric mode, exist at transmission dips and transparency frequencies respectively. The coupling between hybridized dipoles can be understood in terms of the coupled oscillator model and an equivalent circuit model. Further, the EIT response can be tailored through changing the degree of asymmetry. These rich spectral features are benefit from the three dimensional meta-atom design in which the through hole reconstructs the resonance current distributions and magnetic/electric dipole's direction may be rearranged to provide more resonance pattern. The design concept may also be extended to THz and optical spectral. It will lead the application of metamaterials to multifunctional device, such as optical switches, frequency selective devices, optical modulators, and ultrasensitive sensors and slow-light devices [36, 37].

## 5. Funding and Acknowledgments

This work was supported by the Fundamental Research Funds for Central Universities (No.CCNU16A0216) and National Natural Science Foundation of China (No. 41474117).